# Optimum design of tracking bifacial solar farms – A comprehensive global analysis of next-generation PV

M. Tahir Patel[1*], M. Sojib Ahmed[2*], Hassan Imran[3], Nauman Z. Butt[3], M. Ryyan Khan[2], and Muhammad A. Alam[1]

[1]Electrical and Computer Engineering Department, Purdue University, West Lafayette, IN, USA
[2]Department of Electrical and Electronic Engineering, East West University, Dhaka, Bangladesh
[3]Department of Electrical, Lahore University of Management Science, Lahore, Pakistan

*Abstract* – The bifacial gain of East-West vertical and South-facing optimally-tilted bifacial farms are well established. One wonders if bifacial gain (and the associated LCOE) may be further improved by tracking the sun. Tracking bifacial PV has advantages of improved temperature sensitivity, enhanced diffuse and albedo light collection, flattened energy-output, reduced soiling, etc. Monofacial tracking already provides many of these advantages, therefore the relative merits of bifacial tracking are not obvious. In this paper, we use a detailed illumination and temperature-dependent bifacial solar farm model to show that bifacial tracking PV delivers up to 45% energy gain when compared to fixed-tilt bifacial PV near the equator, and ~10% bifacial energy gain over tracking monofacial farm with an albedo of 0.5. An optimum pitch further improves the gain of a tracking bifacial farm. Our results will broaden the scope and understanding of bifacial technology by demonstrating global trends in energy gain for worldwide deployment.

## 1. Introduction

The global energy demand is approaching unprecedented levels with the projected population of 10 billion people by 2050. Protecting the environment while meeting the energy demands will involve a delicate balancing act in the coming decades. It has been suggested that renewable energy resources can satisfy the energy demand with the smallest environmental footprint. Among all renewable energy resources, the direct conversion of solar energy to electricity by photovoltaic technology (PV) is one of the most promising sources of green and sustainable energy with the advantage of limited environmental impact and low maintenance cost.

One way to reduce the levelized cost of energy (LCOE) is to design a photovoltaic (PV) system to produce the maximum energy for a specified area of the solar farm. Among the cost-effective options, one can either increase the total irradiance of fixed-tilt farms using bifacial modules or increase the normal irradiance on monofacial modules by solar-tracking. Both technologies are becoming popular as bifacial modules are expected to have ~40% market share by 2028 [1], whereas solar-tracking is adopted by ~70% of newly installed PV systems since 2015 [2]. The solar-tracking of bifacial PV modules sounds appealing, but the only concern is that whether bifaciality would be beneficial in solar-tracking configuration, as maximizing direct beam will cast a deep shadow on the ground and reduce the contribution from the ground albedo. Fortunately, the recent location-specific case-studies predict the promising potential of bifacial PV farms in solar-tracking configuration [2,3]. For example, the detailed calculation by Egido et al. [4] predicts that a *standalone dual-axis tracking* bifacial PV module (with the ground painted white) has almost 80% higher annual energy yield as compared to its fixed monofacial counterpart. For the same white-paint albedo, a 20% to 90% increase in energy yield was demonstrated across different months for *single-axis tracking* of standalone bifacial PV modules as compared to fixed-tilt bifacial and monofacial modules, respectively. Similarly, Oria and Sala reported a 70% increase in annual energy yield using single-axis tracking of standalone bifacial modules with highly reflective albedo collectors as compared to their fixed monofacial counterparts [5]. Shoukry et al. [6] demonstrated that both bifaciality and tracking have their own unique contributions depending upon the geographical locations with bifaciality providing up to 44% advantage as compared to single-axis tracking that provided 18% energy gain as compared to fixed monofacial counterpart. There are other studies in the literature that demonstrate the promising potential of standalone bifacial and/or tracking PV panels both experimentally [7,8] and computationally [2,9]. There is an emerging consensus that the irradiance incident on a module can be enhanced through bifaciality/tracking.

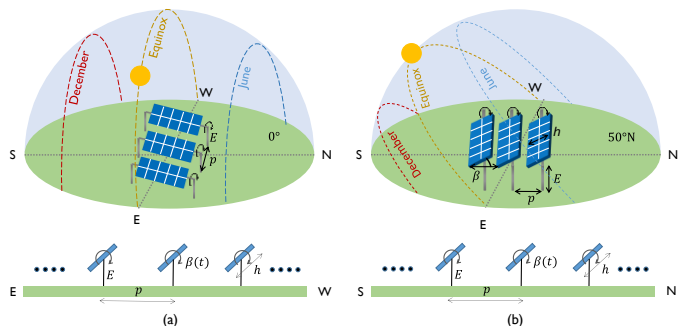

Fig. 1. Schematic of a single-axis tracking bifacial solar PV farm (a) E-W facing (b) N-S facing. The solar path varies with latitude as well as time making one of the designs geographically more suitable.

Despite these promising initial results, one still wonders if temperature-dependent efficiency degradation due to additional solar irradiance would erase any performance gain. Indeed, several research groups have explored the temperature-dependent performance of stand-alone PV modules. By combining the irradiance, light collection, and electro-thermal models, several research groups have demonstrated that the bifacial gain and corresponding reduction in LCOE depend on the geographical location and one must optimize the module tilt and row-separation to maximize the bifacial gain [10,11].

Along with standalone PV modules, several research groups investigated monofacial/bifacial PV farms both in fixed and/or mobile configurations. Stefano et al., [12] demonstrated that the bifaciality of single-axis mobile PV arrays could enhance their performance by 12%. Stein et al., [13] reported ~10% bifacial gain for single-axis mobile PV systems. By using their mathematical model, Janssen et al., demonstrated that single-axis tracking of monofacial and bifacial PV arrays could increase energy yield up to 15% and 26% respectively as compared to standard monofacial fixed-tilt system. A number of studies also reported the temperature-dependent performance of tracking PV systems based on theoretical analysis or simulated results [14–16]. Recently, based on their optical-electrical-thermal model, Patel et al., [17] showed that temperature-dependent efficiency could change LCOE of fixed-tilt bifacial solar farms by -10% to 15% worldwide for different module technologies. Using a collection of solar farms in Brazil, Verissimo et al. suggest that country-specific tracking results show solar-farm topology can have a significant impact on the energy yield and LCOE of the system [18]. Rodriguez-Gallegos et al. [19] have also shown that the single-axis tracking bifacial PV system exhibit minimum LCOE worldwide, even though dual-axis tracking bifacial PV systems produce higher energy. A dual-axis system, however, has higher LCOE because its higher energy yield is partially counterbalanced by its higher initial, operational, and maintenance costs as compared to that for the single-axis system. After all, the single-axis system is simpler to design and operate both in E/W facing orientation to track the sun daily and in N/S facing orientation to track the sun over the seasons. Without taking into account location-specific land cost, the study [19] showed that single-axis bifacial PV farms exhibit a 16% decrease in LCOE for locations at latitudes within a range of ±60°. These results further demand a careful evaluation of global-scale LCOE for single-axis tracking bifacial PV farms by considering the site-specific land costs. Indeed, as discussed in Ref. [20], the land-cost dramatically changes the optimum spacing, and the energy yield of a solar farm.

This work is the culmination of our series of studies on *standalone* and farm-level bifacial PV design considerations. We performed worldwide analyses investigating the optimum design parameters for energy maximization of a single bifacial PV module [10], a vertical bifacial PV farm [21], a ground-sculpted bifacial PV farm [22], and cost minimization of a fixed-tilt bifacial PV farm [17,20]. The purpose of this paper is to investigate the worldwide performance of bifacial single-axis tracking PV farms with fixed-tilt bifacial PV farms. We compare the performance comparison of two different tracking algorithms termed as *tracking the sun* (TS) and *tracking the best orientation* (TBO) [23]. By using a self-consistent optical-electrical-thermal model, we address the following key questions: (i) How do the energy yields compare between east/west (E/W) facing and north/south (N/S) facing bifacial single-axis tracking solar farms? (ii) How does the site-specific climate affect the performance of these farms? (iii) How do the location-dependent land costs affect the optimum design and LCOE of these farms? and (iv) What is the optimum row-spacing (pitch) for minimum LCOE?

This paper is divided into four sections. Section II describes the modeling approach. Section III discusses the results and the conclusions are provided in Section IV.

## 2. MODELING FRAMEWORK

To model a tracking solar PV farm, one must generalize and integrate several sub-models developed in previous studies [10,17,20–22]. We begin by integrating the irradiance model and a tracking algorithm to find the intensity of direct light falling on the solar modules. Next, we used a light collection model with view factors and masking formulae to estimate the diffuse and albedo light collection. Finally, the thermal model combined with the PV energy (electrical) model produces the total energy generated by the solar PV farm. A flow diagram in Fig. 2 demonstrates the complexity of a self-consistent model necessary to predict the performance of a tacking solar farm.

### 2.1 PV farm power and energy model

As shown in Fig. 1, we model a solar farm with the following design parameters: azimuth angle ($\gamma_A$) from the North; panel height ($h$); panel angle ($\beta$); elevation from the ground ($E$); pitch ($p$); and ground albedo ($R_A$).

First, we use the ***irradiance model*** and estimate the ground illumination for a particular time of the day at a location specified by its latitude and longitude. We follow the modeling described in [17,20,21]. We use the PV library from Sandia National Laboratory to calculate the Sun's trajectory (zenith ($\theta_Z$) and azimuth angle ($A$)) and the irradiance [21]. The Global Horizontal Irradiance (GHI or $I_{GHI}$) is ideally given by Haurwitz clear sky model [24,25]. Renormalizing this irradiance based on the NASA Surface meteorology and Solar Energy database [26] yields the local variation in GHI. The GHI is then split into direct light (DNI or $I_{DNI}$) and diffuse light (DHI or $I_{DHI}$) using Orgill and Hollands model [27]. Perez model [28] is further used to find circumsolar and isotropic components of diffuse light.

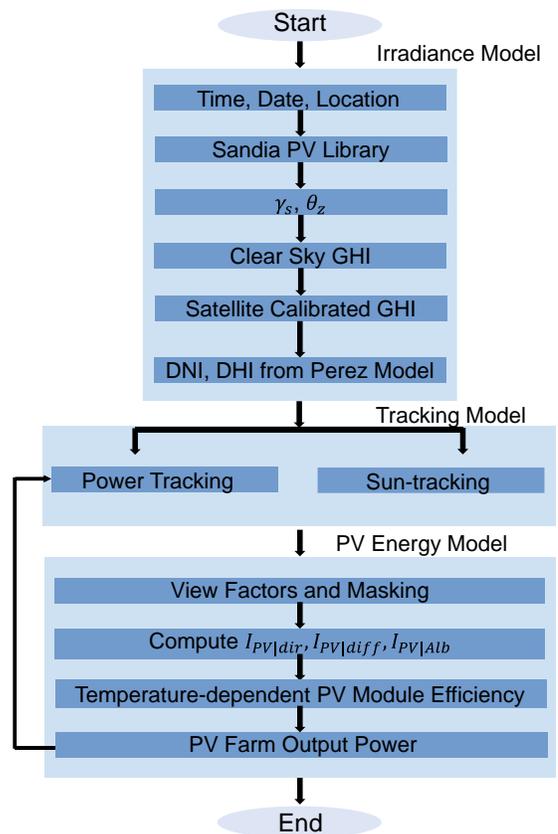

Fig. 2 The simulation workflow integrates irradiance, light collection, and tracking models to calculate the energy yield of a bifacial tracking solar farm.

TABLE I
EQUATIONS ASSOCIATED WITH THE OPTICAL MODEL: IRRADIANCE MODELING AND LIGHT COLLECTION

| | |
|---|---|
| (1) $I_{GHI} = I_{DNI} \times \cos(\theta_Z) + I_{DHI}$ | (6) $I_{PV:Alb.diff}^{F,Panel}(l) = I_{Gnd:DHI} \times R_A \times F_{dl-gnd}(E,l) \times \eta_{diff}$ |
| (2) $I_{PV:DNI}^{F,Farm} = I_{DNI} \cos\theta_F (1 - R(\theta_F))\eta_{dir}$ | (7) $I_{PV:Alb.diff}^{F,Farm} = 1/h \int_0^h I_{PV:Alb.diff}^{F,Panel}(E,l)\, dl$ |
| (3) $I_{PV:DHI}^{F,Farm} = I_{DHI}\, \eta_{diff}/h \int_0^h \frac{1}{2}(1 + \cos(\psi(E,l) + \beta))\, dl$ | (8) $I_{PV:Alb}^{Farm} = I_{PV:Alb.dir}^{Farm} + I_{PV:Alb.diff}^{Farm}$ |
| (4) $I_{PV:Alb.dir}^{F,Panel}(l) = I_{Gnd:DNI} \times R_A \times F_{dl-gnd}(E,l) \times \eta_{diff}$ | (9) $I_{PV(T)}^{Total} = I_{PV:DNI}^{Farm} + I_{PV:DHI}^{Farm} + I_{PV:Alb}^{Farm}$ |
| (5) $I_{PV:Alb.dir}^{F,Farm}(l) = 1/h \int_0^h I_{PV:Alb.dir}^{F,Panel}(E,l)\, dl$ | (10) $YY_T(p,\beta,h,E,\gamma_A,R_A) = \int_0^1 I_{PV(T)}^{Total}(p,\beta,h,E,\gamma_A,R_A)\, dY$ |

TABLE II
EQUATIONS ASSOCIATED WITH THE THERMAL MODEL

| | |
|---|---|
| (11) $\eta(T_{cell}) = \eta_{STC}(1 - TC(T_{cell} - T_{STC}))$ | (14) $T_{cell} = T_{amb} + (P_{POA} - \eta P_{POA} - \gamma P_{POA}(subband)) * \frac{F}{1000}$ |
| (12) $T_{cell} = T_{amb} + P_{POA} \times e^{a+(b \times WS)} + \Delta T \times \frac{P_{POA}}{1000}$ | (15) $T_{cell} = T_{amb} + \left(P_{POA_{top}} - \eta_{top} \times P_{POA_{top}} + P_{POA_{bot}} - \eta_{bot} \times P_{POA_{bot}} - P_{POA}(subband)\right) * \frac{F}{1000}$ |
| (13) $T_{cell} = T_{amb} + c_T \frac{\tau\alpha}{u_L} P_{POA}\left(1 - \frac{\eta(T_{cell})}{\tau\alpha}\right)$ | (16) $P_{POA} = I_{PV:DNI}^{Farm}/\eta_{dir} + (I_{PV:DHI}^{Farm} + I_{PV:Alb}^{Farm})/\eta_{diff}$ |

Second, we quantify the amount of **light collected by the solar panels** installed at that location. The panels have height $h$, tilted at an angle $\beta$, separated by pitch (or period) $p$, and are oriented at an array azimuth angle $\gamma_A = 180°$ (i.e., south-facing panels) for farms in the northern hemisphere and $\gamma_A = 0°$ (i.e., north-facing panels) for farms in the southern hemisphere or $\gamma_A = 90°$ for panels facing E-W direction. The collection of light on panels from the three components of irradiance (i.e., direct, diffuse, and albedo) are formulated separately and analyzed accordingly. Our approach to model the collection of light involves the view-factor calculation described in [17,20]. The corresponding equations are summarized in Table 1.

Third, we incorporate the **temperature-dependent efficiency model** which require the collected light intensity, efficiency at STC, and ambient temperature as inputs and yield the effective efficiency as output [17]. The temperature-dependent efficiency loss involves a complex interplay of increased absorption due to bandgap reduction vs. increased dark-current and reduced mobility. At practical illumination intensity, it is well-known that the efficiency $\eta(T)$ scales linearly with the module/cell temperature, see Eq. 1 in Table II where the rate/slope of degradation is given by the absolute temperature coefficient ($TC$) [29,30].

Finally, we find **the daily, monthly, and yearly energy-output** of the farm. Using the equations (2), (3), and (8) in Table-I, we arrive at Eq. (9) to find the time-varying spatially distributed light collection on the panels. This information combined with the thermal model (summarized in Table-II) is used in the circuit model to find the equivalent power generation. To estimate energy output, we integrate the power generated over the desired period of time. We define the energy yield per pitch of a farm over one year as yearly yield ($YY$) given by Eq. (10).

### 2.2 Tracking model

In a PV tracking system, instead of a fixed panel tilt, the tilt angle $\beta(t)$ varies with time. There are two types of tracking algorithms: (a) **sun-tracking** orients the modules to maximize the direct light, and (b) **power-tracking** orients the modules to maximize power or energy.

As the name suggests, a **sun-tracking** system minimizes the angle between the sun and the panel. To compute the time-dependent tilt angle of E/W facing modules by the **sun-tracking algorithm**, first, we compute the angle of incidence ($\theta_{AOI}$) between the direct light (DNI) and the normal of the module face. Solar tracking (TS) minimizes the angle of incidence towards 0° so that the face of the solar panel follows the sun (direct light). Fig. 3 shows the hourly variation of the tilt angle $\beta$ for TS for June and December at Washington D.C. (38.9° N, 77.0° W) and Dubai (24.5° N, 55.5° E) in the northern hemisphere.

For E/W tracking, the panels follow the sun throughout the day, and thus the tilt angle $\beta(t)$ goes from $+90°$ (East-facing) to $-90°$ (West-facing) from sunrise to sunset. For N/S tracking, during summer (June) around sunrise, $\beta(t)$ is negative i.e., North-facing panels with $\gamma_A = 0°$ becomes close to 0° at solar noon and goes back to negative during sunset. This is because the sun's path, from sunrise to sunset, lies almost entirely in the North except near solar noon where the sun could be in the south of the panels. Whereas, during winter (December), $\beta(t)$ remains always positive i.e., South-facing panels with $\gamma_A =$ $180°$ because the sun's path is entirely in the South. Further, the panels face the south-inclined sun normally at solar noon where the $\beta(t)$ is equal to the zenith angle of the sun. Thus, the sign of tilt angle (and the array azimuth angle) varies with the sun's path according to the latitude of a geographical location throughout the year. The days in December are shorter than days in June, which is evident from the sunrise and sunset points. Moreover, $\theta_{AOI} = 0°$ is not achievable for both months, thus, the algorithm minimizes $\theta_{AOI}$ for the closest normal incidence. Finally, the initial angle being $\pm 90°$ is impractical for a solar farm because in this scenario the first row of panels will cast a row-to-row shadow on all the other panels.

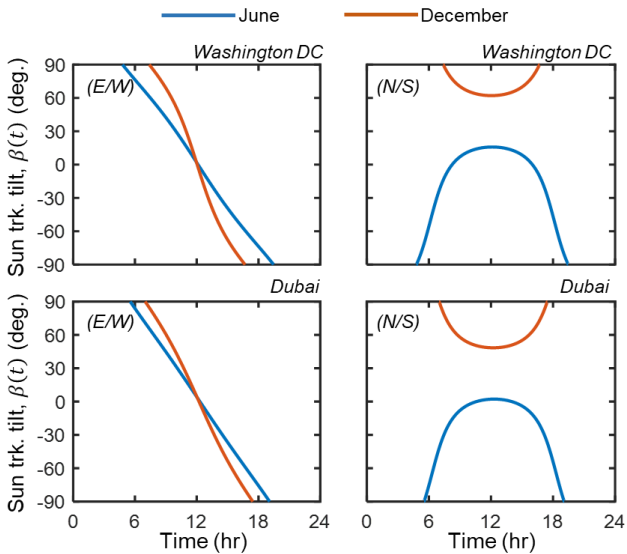

Fig. 3 Using sun-tracking algorithm, we plot the tilt angle as a function of time for EW facing tracking PV modules (left column) and NS facing tracking PV modules (right column) for June (Blue) and December (Red) at Washington, D.C., $38.9°$ N, $77.0°$ W (top row) and Dubai, $24.5°$ N, $55.5°$ E (bottom row).

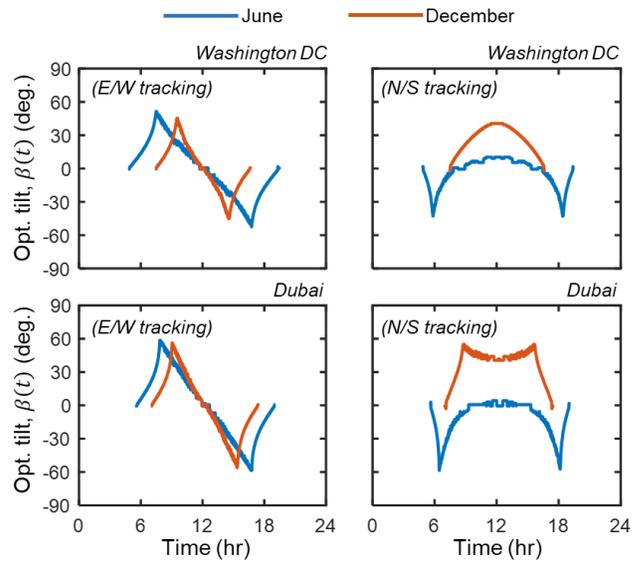

Fig. 4 Using power-tracking algorithm, we plot the tilt angle as a function of time for EW facing tracking PV modules (left column) and NS facing tracking PV modules (right column) for June (Blue) and December (Red) at Washington, D.C. (top row) and Dubai (bottom row).

The second approach involving **power-tracking** optimizes the time-dependent tilt angles to achieve maximum power at each time-step—this will also maximize daily energy. The optimization is governed by the combined effects of (a) relative contributions of instantaneous DHI and DNI, (b) position of the sun, and (c) panel-to-panel shading. This power-tracking or the best orientation (TBO) algorithm is targeted towards maximizing output, not the light collection. Thus, during the early and late part of the day, the system will compromise light collection to avoid shading loss and maximize power output.

Fig. 4 shows the time-dependent optimized tracking angles $\beta(t)$ in June and December for Washington D.C. and Dubai. During the early and late parts of the day, the irradiation is mostly sky-diffuse light. At these times, in contrast to the TS algorithm, the panel optimally faces the sky (i.e., $\beta(t) \approx 0$) for both E/W and N/S tracking configurations. As the day progresses, the fractional contribution of DNI increases, and the $\beta(t)$ will compete between facing the sky and the sun. This is prominently observed for optimal tilt for E/W tracking panels in Fig. 4. After a certain time, e.g., 7 am in June, the panels closely follow the sun. In general, even when the DNI fraction is higher, the $\beta(t)$ is shallower than sun-tracking angles to improve the collection of DHI. The *N*-shape profile of $\beta(t)$ in E/W tracking is narrower in winter seasons due to shorter days. For N/S tracking, $\beta(t)$ is larger in December compared to June due to the more tilted sun-path.

To compare the tracking algorithms in terms of insolation collection, Rodriguez-Gallegos et al. [23] performed a PV tracker performance analysis using these two algorithms – tracking the sun (TS) and the best orientation (TBO) – for 3 different configurations, namely, horizontal single-axis tracker (HSAT), tilted single-axis tracker (TSAT), and dual-axis tracker (2T). They found that TBO shows a higher insolation collection compared to TS but the difference is below 1.8% for |latitude| $< 60°$ and 3.3%, 7.1%, and 2.9% for HSAT, TSAT, and 2T, respectively.

A few points regarding practical tracking algorithms are noteworthy. First, the power-tracking or TBO algorithm requires the panels to back-track at least twice during the day (observe the peaks and troughs in Fig. 4). This may cause additional mechanical wear and tear of moving parts and electrical issues with the motor. In contrast, the sun-tracking algorithm continuously follows the sun during the day which avoids backtracking and related reliability issues, if any. Second, the maximum value of the tilt angle for tracking the panels is $60°$ to avoid rotational load on the axis and avoid mutual shading between the rows of panels at higher tilt angles. Third, in practice, the panels are kept at an initial tilt of $60°$ (not horizontal at $0°$) to reduce dew-enhanced soiling that would reduce the output of the solar panels.

In the following section, we will discuss the energy yield of solar farms based on the power-tracking algorithm.

## 3. POWER AND ENERGY YIELDS AT SPECIFIC LOCATIONS

In this section, we will summarize the modeling and simulation results to energy yield and then discuss LCOE-

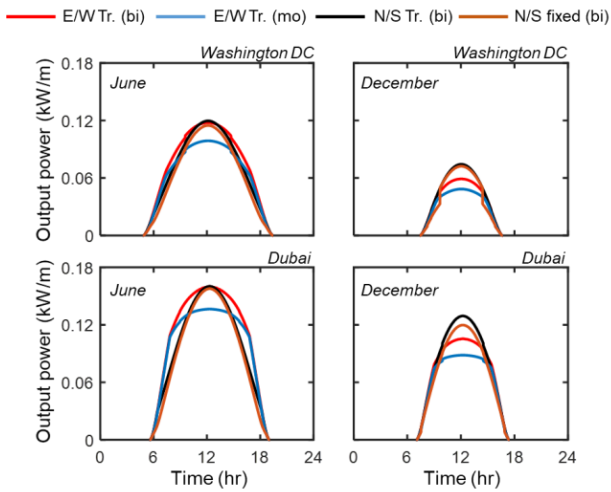

Fig. 5 Daily output power for Washington D.C. and Dubai during summer months (June) and winter months (December). E/W tracking performs better in summer months and lower latitudes while N/S tracking has a higher output during winter months and higher latitudes.

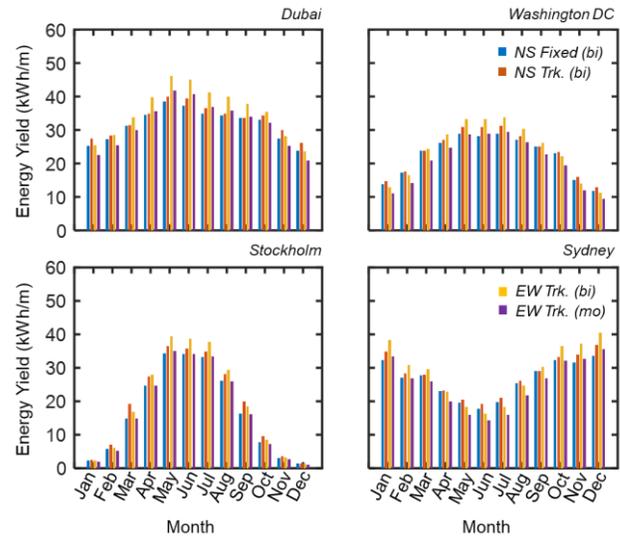

Fig. 6 Monthly energy yield for various farm designs, namely, N/S Fixed tilt (Blue), N/S Tracking (Red), E/W Bifacial Tracking (Yellow), and E/W Monofacial Tracking (Purple).

optimized solar farms. For an intuitive understanding of the results, we first compare the monthly energy output for various farm designs at four locations – Dubai (24.5° N, 55.5° E), Washington, D.C.(38.9° N, −77.0° W), Stockholm (59.3° N, 17.8° E), and Sydney (33.8° S, 150.6° E). Next, we present the global energy yield maps for fixed-tilt, East-West (E/W), and North-South (N/S) tracking farm designs. By default, we assume an east-west (E/W) tracking bifacial farm with a pitch over height, $p/h = 2$ and elevation above the ground, $E = 1m$ (we will discuss pitch and height-optimized design later in section 4). We then compare this default design with various other PV farm configurations, namely, fixed-tilt bifacial farm, E/W tracking monofacial farm, and N/S tracking bifacial farm. Finally, we will present the effect of varying pitch on the yearly energy yield and LCOE* for the default EW tracking PV farm design.

*3.1 Power output*

*E/W tracking vs. N/S Fixed-tilt Systems.* The power outputs of the PV farm using the power-tracking algorithm along with N/S fixed-tilt systems are shown in Fig. 5. We find that the energy yields of E/W tracking bifacial farms are always higher than that of E/W tracking monofacial farms for Dubai and Washington, D.C. In June (for both Washington DC and Dubai), E/W tracking yields higher energy compared to N/S systems slightly before and afternoon, resulting in overall higher output from E/W tracking PV. In December, as the sun-path is more tilted, the E/W systems cannot adequately follow the sun ($\theta_{AOI}$ is high even at noon). Thus, the N/S tracking and fixed-tilt systems give more output than E/W tracking systems in December. Although the net gain depends on the location, the counterbalancing energy yield for summer and winter holds for any location, as discussed in the next section.

*N/S tracking vs. N/S Fixed-tilt Systems.* The N/S tracking panels have a higher output than N/S fixed-tilt panels as it can avoid shading in mornings and evenings and correct for panel tilt for seasonal sun-path variations. For example, in Fig. 5(b) there is a clear effect of partial shading in Washington DC in December for N/S fixed panels (observe the lowered output before 9 am or after 3 pm).

*3.2 Monthly/Seasonal energy yield variation*

The monthly energy yield for various PV module configurations is shown in Fig. 6. Various locations in both the northern and the southern hemispheres are explored here. During summer months in the respective hemispheres, E/W bifacial solar tracking scheme performs best, whereas, in the other months, N/S bifacial solar tracking scheme performs better. One of the important points to note is that N/S bifacial fixed-tilt performs best during Nov. to Jan. in Stockholm. Here, although higher irradiance is collected at bifacial modules for N/S solar tracking configuration, higher module temperature lowers its efficiency as compared to its fixed-tilt counterpart. Moreover, the average monthly energy gains for Dubai, Washington, D.C., Stockholm, and Sydney, respectively are as follows -

(i) E/W tracking vs. N/S tracking: 5.75%, 0.16%, − 3.00%, and 1.08%.
(ii) E/W tracking bifacial vs E/W tracking monofacial: ~11%, ~15%, ~18%, and ~13%.
(iii) N/S tracking vs. N/S fixed tilt: ~6%, ~5%, ~14%, and ~6%.

The yearly minimum and maximum energy gains are shown in Fig. S4. The minimum is observed in the winter months and maximum in the summer months of the respective hemisphere. In terms of energy yield, it is apparent from these results that E/W tracking would be better near the equator and N/S tracking would be beneficial for higher latitudes.

We have so far used a fixed $p/h = 2$ for the daily and monthly energy calculations, however, the same $p/h$ would not generate the maximum energy for all the farm designs due to the difference in row-to-row shading throughout the day. Thus, for energy maximization and cost minimization, we need to find

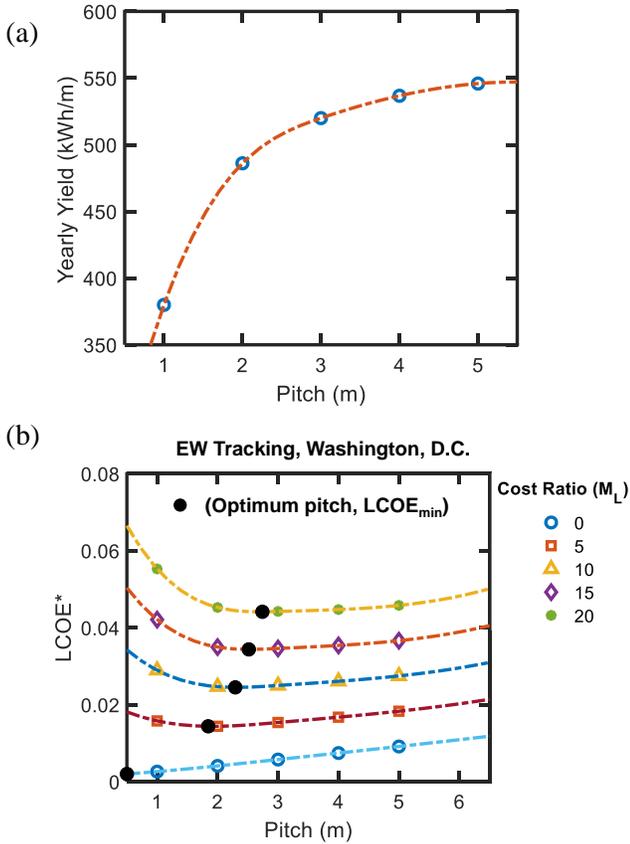

Fig. 7 (a) An increasing trend in yearly energy yield with increasing pitch. The yearly energy yield saturates at higher values of pitch. (b) For a give value of cost ratio ($M_L$), an optimum pitch provides a minimum LCOE*. The optimum pitch increases with increase in module to land cost ratio ($M_L$).

the optimum pitch over height ratio ($p/h$) for specific farm design, as discussed in the next section.

### 3.3 OPTIMUM $p/h$, YY, AND LCOE-MIN

The Pitch of a solar farm – defined as the spacing between the rows of the PV panels – is an important design parameter. A farm is optimized by first quantifying how yearly energy yield and LCOE depend on the pitch of a solar farm.

For the daily power output and monthly energy calculations in the previous sections, we assumed the pitch over height ratio $p/h = 2$, for all the locations. In practice, higher $p/h$ improves the energy yield (per panel) due to an increase in albedo light collections and reduction in mutual shading between the rows of solar panels, see Fig. 7(a). However, once land-cost is brought into consideration (in other words, when the land is limited), then an optimum value of $p/h$ yields the minimum LCOE. This optimum $p/h$ is depicted in Fig. 7(b) for an E/W-tracking system at Washington, D.C. We find that the optimum p/h increases with increasing module-to-land cost-ratio ($M_L$). For a given module-related cost, a decrease in cost-ratio implies an increase in the land-related costs, leading to a reduction in the optimum pitch. Moreover, the optimum pitch and the associated minimum value of LCOE varies with the geographical location.

### 4. GLOBAL-SCALE MERITS OF VARIOUS FARM DESIGNS

Given the understanding of power and energy-yield at the four cities across the world, now we are ready to compare the technologies across the globe. As discussed in Ref. [20], however, we must optimize the row-spacing depending on the LCOE-considerations.

In recent years, the $M_L$ has ranged from 9 to 15 [20,31]. Therefore, the optimum p/h for minimum LCOE lies between 2 and 3 for most locations around the world, see Fig. 8(c). For $M_L = 12$, the global yearly energy yield and minimum LCOE* at optimum p/h are shown in Fig. 8. (This result led to the default choice of p/h = 2 in Sec. 3.) Nonetheless, it is preferred that prior to the PV farm deployment, we optimize the farm design through an analysis of LCOE (proportional to LCOE*) variation with the pitch for each farm site around the world.

### 4.1 Energy Yield of E-W Tracking Bifacial Solar Farms

Fig. 8(a) shows the yearly energy yield of an E/W tracking bifacial PV farm with an optimum pitch over height for $M_L = 12$. The yearly energy output per unit panel height reaches approximately 600 kWh/m for geographical locations close to the equator and deserts across the globe. These locations receive the highest fractions of direct light and thus, a tracking system benefits the most at these places. However, for locations with high diffuse light fraction, i.e., Beijing, Northern Europe, and North America, the yearly yield is significantly lower, $\sim 200 - 300$ kWh/m. The E/W tracking PV farms are not advantageous in these regions. The yearly energy yield in Fig. 8 will be taken as a reference for the comparisons performed in the following sections.

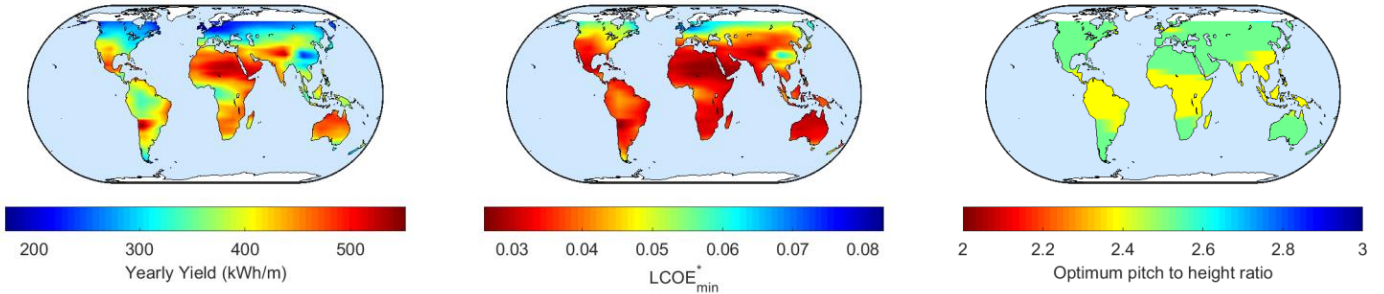

Fig. 8 Global trends in yearly energy yield, LCOE*, and optimum pitch/height of an elevated bifacial farm ($E = 1m$) for $M_L = 12$.

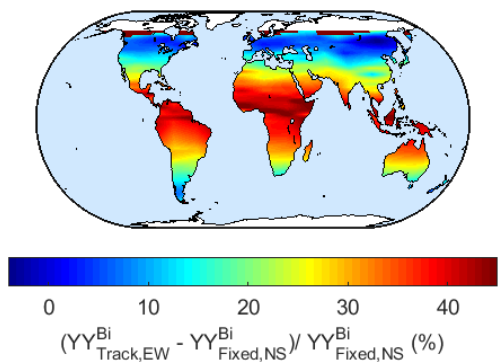

Fig. 9 Percentage gain in yearly energy yield in East-West tracking bifacial PV farm compared to North-South facing fixed-tilt bifacial farm. The energy gain for E/W tracking is highest near the equator and in the deserts around the world whereas fixed-tilt performs slightly better at very high latitudes.

*4.2 Tracking PV farm vs. fixed-tilt bifacial farm*

Let us now determine if a single-axis tracking PV farm should be a preferable choice of deployment as compared to a fixed-tilt farm at any specific location around the world. The energy surplus/deficit of a tracking PV farm over a fixed-tilt PV farm depends on the geographical location, see Fig. 9. The E/W tracking bifacial PV farm outperforms the N/S fixed-tilt bifacial PV farm by ~45% near the equator and at lower latitudes. The highest energy gain is observed in desert regions, e.g. Sahara, Atacama, Australian, Kalahari, etc. These regions receive a high fraction of direct light irradiance and EW tracking PV relies on tracing the direct light, thus leading to increased energy gain. However, for sites with low direct light fraction and in turn high diffuse fractions, e.g., |latitudes| > 45° and Northeast China, the fixed-tilt bifacial farm is comparable or slightly better in energy gain. Given the additional land-area needed, higher initial and O&M cost, tracking may not be suitable for these locations.

Moreover, a comparison between E/W tracking monofacial farm and N/S fixed-tilt bifacial farm shows that tracking monofacial farm is advantageous only for locations with high fractions of direct light i.e., |latitude| < 30°, whereas N/S fixed-tilt bifacial farm performs better for higher latitudes and for regions where the diffuse light fraction is relatively higher. Fig. S1 clearly displays this trend.

Overall, it will come as no surprise that tracking PV farms are advantageous for locations with a high direct light fraction while fixed-tilt bifacial farms perform better at sites with the high diffuse light fraction. Our contribution is to quantify the energy gain associated with various configurations of optimized solar farms.

*4.3 E/W vs. N/S Tracking bifacial farm design*

Next, we explore a conventional E/W single-axis tracking PV farm design with an N/S single-axis tracking one. In contrast to the hourly tracking of direct sunlight in an E/W tracking PV, an N/S tracking farm tracks the seasonal change in sun path (solar elevation/zenith angle) and varies comparatively less during the day. Since N/S fixed-tilt bifacial farm generates more energy output as compared to an E/W tracking farm at higher latitudes, we expect an N/S tracking farm to perform even better than a fixed-tilt bifacial farm.

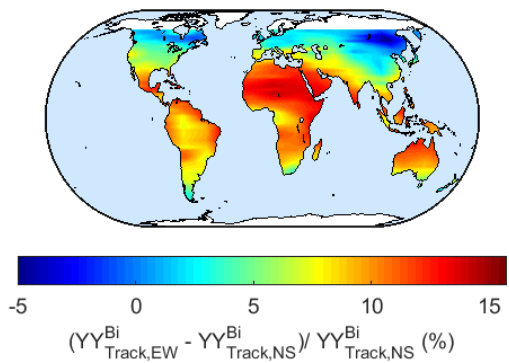

Fig. 10 Percentage gain in yearly energy yield in East-West tracking bifacial PV farm compared to North-South tracking bifacial PV farm. The energy gain for E/W tracking is highest near the equator and in the deserts around the world whereas N/S tracking performs better at very high latitudes.

Fig. 10 displays a worldwide comparison between an E/W tracking farm and an N/S tracking farm. As expected, we observe that N/S tracking farm to achieve higher energy output than E/W tracking farm for locations with high diffuse light fraction, e.g., Northern Canada and Siberia. Moreover, in comparison to E/W tracking farm, the (negative) percentage

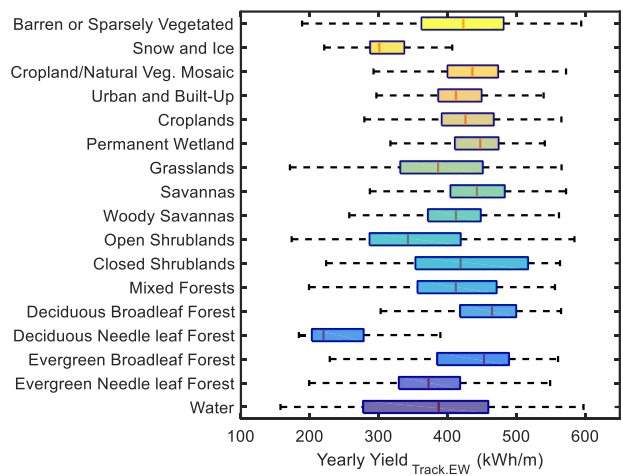

(a)

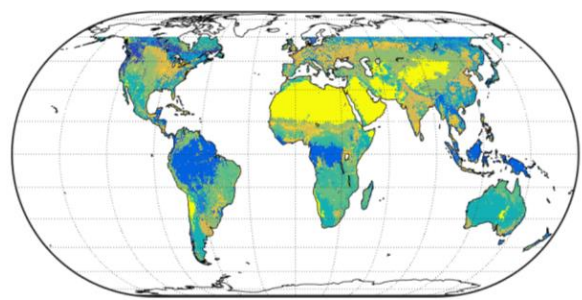

(b)

Fig. 11 (a) Yearly energy yield of E/W tracking bifacial PV farm over various types of land. (b) A map of the land types corresponding to the y-axis in (a). "0" corresponds to "Water" while "16" refers to "Barren or sparsely vegetated" land type.

yearly energy gain is higher for an N/S tracking farm (Fig. 10) than an N/S fixed-tilt bifacial PV farm (Fig. 9) for higher latitudes.

Thus, with ~15% energy gain, E/W single-axis tracking PV farm is a better design for locations with |latitude| < 50° and regions with a high direct light fraction. An N/S single-axis tracking farm on the other hand shows ~5% higher energy yield than the E/W tracking design for sites with |latitude| > 50° and those with a high diffuse light fraction.

*4.4 Does Land-type Matter for Energy Yield?*

In the previous sections, we have analyzed the global energy yields for various tracking and fixed-tilt PV configurations. However, it is interesting to examine these solar farm designs in the context of the type of land of the installation site. This investigation may be especially useful for the new range of agro-photovoltaic (agrivoltaics) applications around the world. For example, it has been recently claimed that croplands in moderate climatic conditions are preferred locations for solar farms as well [32]. Fig. 11 summarizes our result of the analysis succinctly. We consider 17 types of land as shown on the y-axis of Fig. 11(a) – the data was collected from Ref. [33]. These land types are also marked on the world map in Fig. 11(b), where "16" corresponds to "Barren or Sparsely Vegetated" land, and "0" refers to "Water". We observe that, on an average, maximum energy yield is expected for the land types "2" (Evergreen Broadleaf Forest) and "4" (Deciduous Broadleaf Forest). However, these land types (2 and 4) are rich in biodiversity and we should avoid installing PV farms here to preserve ecological balance. "Savannas" (9) and "Barren lands/Deserts" (16) also show good potential in terms of yearly energy yield. Land type "3" (Deciduous Needleleaf Forest) and "15" (Snow and Ice) are the least favorite for tracking EW bifacial PV. In summary, we can infer from Fig. 11 that installing EW tracking bifacial PV farms in parts of Deserts, and Savannas are highly viable. The regions of croplands and grasslands have medium potential for energy yield (by E/W tracking) can thus be combined with agricultural usage and animal husbandry.

**5. SUMMARY AND CONCLUSION**

In this paper, we have investigated the energy yield and comparative performance of various solar farm design configurations such as single-axis tracking vs. fixed-tilt and bifacial vs. monofacial. Our modeling framework combines the irradiance model, light collection model, temperature-dependent efficiency model, and a single-axis tracking model to estimate the energy yield of a solar farm. The tracking algorithm constraints the panel tilt angle to vary such that the angle of incidence is 0° i.e., the direct light falls normally on the panel. These models are simulated for locations worldwide that finally output global maps which quantify the percentage change in energy yield while comparing various farm designs. The analysis of this paper leads to the following key conclusions:

- An E/W single-axis tracking bifacial PV farm generates up to ~45% higher yearly energy yield than an N/S facing fixed-tilt bifacial PV farm for locations at |latitude| < 30°, as shown in Fig. 9. A higher fraction of direct light leads to higher energy gain. An E/W tracking monofacial farm vs. fixed-tilt bifacial farm shows similar global trends with maximum energy gain reaching ~10% near the equator (Fig. S1). However, fixed-tilt bifacial farm outperforms tracking monofacial farm by ~5 − 15% for |latitude| > 30° (Fig. S2).
- As shown in Fig. 10, an E/W single-axis tracking bifacial PV farm tracks the hourly movement of the Sun whereas an N/S single-axis tracking bifacial farm mostly tracks the seasonal/monthly movement of the Sun. The former provides up to ~15% more energy for locations close to the equator (|Latitude| < 50°) and the latter generates up to 5% for locations close to the poles (|Latitude| > 50°).
- Monthly/Seasonal variation in energy yield shows that the E/W tracking produces higher energy output in summer months whereas N/S tracking is favorable in winter months due to a southward inclination of the solar path in the Northern hemisphere and vice versa for the Southern hemisphere, as seen in Fig. 6.
- On an average, maximum PV energy yield is achieved at

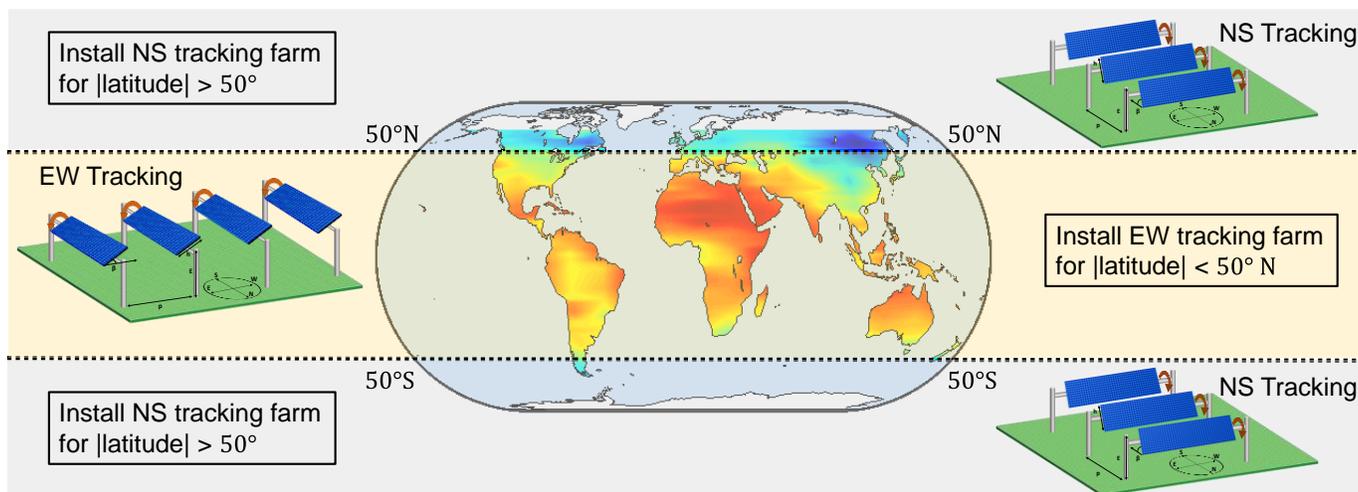

Fig. 12 In conclusion, EW tracking bifacial farms outperform NS tracking bifacial farms for |latitude| < 50° and vice-versa for |latitude| > 50°.

"Evergreen and Deciduous Broadleaf Forests" land types. However, to preserve the biodiversity at these forests, it would be ideal to install E/W bifacial tracking PV on the next best land type for high energy yield, namely, "Deserts" and "Savannas", as illustrated in Fig. 11.

- An optimum pitch over height ratio yields minimum LCOE of an E/W tracking farm. The value of optimum p/h increases with the module-to-land cost-ratio ($M_L$) and varies with the geographical location of the farm. Whereas the yearly energy of the farm increases monotonically with p/h and saturates for very high values of pitch, as shown in Fig. 7. For a typical value of $M_L = 12$, the optimum p/h lies between 2 and 3, as seen globally in Fig 8.

Fig. 12 summarizes our key conclusion that an E/W single-axis tracking bifacial PV farm is the best PV farm design for most regions (|Latitude| < 50°) around the world. The energy gain with respect to a fixed-tilt bifacial or an N/S tracking farm varies according to the incident direct light fraction and the solar path at the geographical location of the solar PV farm. Moreover, for minimizing LCOE of a farm an optimum pitch can be used according to the estimated essential module to land cost ratio ($M_L$) for the deployment site. Thus, in terms of energy maximization, a bifacial tracking PV would be a worthwhile farm technology in most locations of the world.

ACKNOWLEDGMENT

M. Tahir Patel and M. Sojib Ahmed contributed equally to this work. This work is supported by the National Science Foundation under Grant No. #1724728. The authors gratefully acknowledge discussions with Dr. Chris Deline at National Renewable Research Laboratory and Dr. Jim Joseph John from Dubai Electricity and Water Authority for answering questions about practical tracking solar farms. The authors thank Prof. Rakesh Agrawal and Prof. Peter Bermel for discussions on Agro-photovoltaics and for providing the high-performance computational resources needed for this work. Finally, the data and maps will be available in the explorable form based on the work by Dr. Ann Christine Catlin and Chandima Nadungodage.

APPENDIX

*S1: Monofacial E/W tracking vs. N/S fixed bifacial*

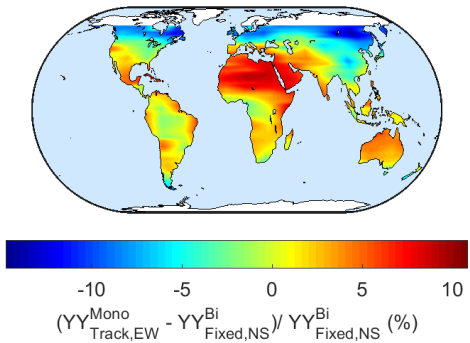

Fig. S1 Percentage change in yearly energy yield between an EW tracking monofacial PV farm and a NS facing fixed-tilt bifacial farm.

*S2: Bifacial gain for EW tracking PV farm*

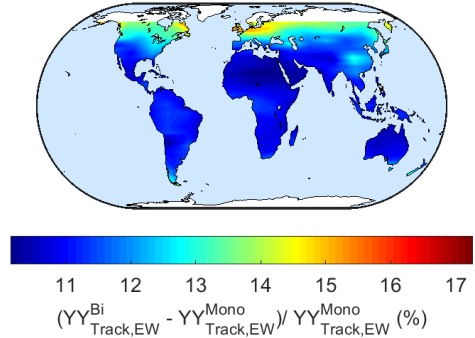

Fig. S2 Percentage change in yearly energy yield between an EW tracking bifacial PV farm and an EW tracking monofacial farm.

*S3: E/W Tracking PV temperature dependence*

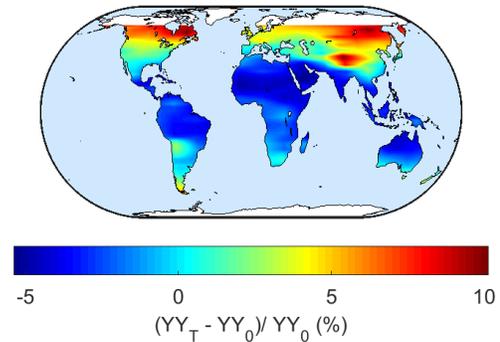

Fig. S3 Difference in yearly yield estimation for E/W tracking bifacial PV farm with and without temperature-dependent efficiency model.

*S4: Monthly energy gain between various PV farms*

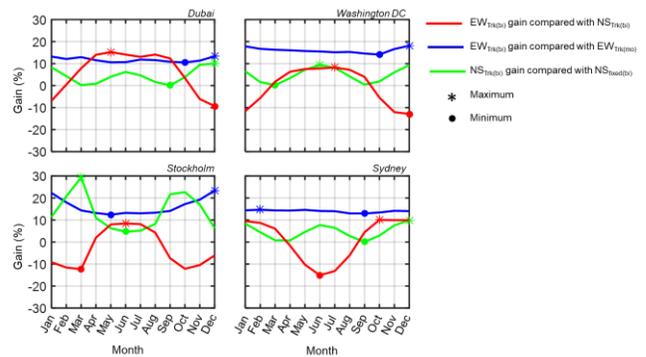

Fig. S4 Difference in monthly energy yield estimation between combinations of E/W tracking, N/S tracking, and N/S fixed tilt bifacial PV farms. The maximum values are denoted by an Asterix (*) and minimum by a dot (.).

Table III. Glossary of the symbols used in this paper

| Parameters | Definition |
|---|---|
| TC | Temperature Coefficient (%/°C) |
| $T_a$ | Ambient Temperature (°C) |
| $T_M$ | Module Temperature (°C) |
| $P_{in}/P_{out}$ | Input Optical Power Collected by the Panel / Output Electrical Power (W/m$^2$) |
| $\eta$ | Power conversion efficiency (%) |
| I/Irradiance | Input optical power over the plane of array (W/m$^2$) |
| a,b | Fitting parameters in the King's model (no unit, s/m) |
| WS | Wind Speed (m/s) |
| $c_T$ | Correction term used for the daily average temperature data |
| $\tau$ | Transmittance of glazing (no unit) |
| $\alpha$ | Absorbed fraction (no unit) |
| $u_L$ | Heat loss coefficient (W/m$^2$K) |
| $\gamma$ | Coefficient of sub-band power contribution to heating (No unit) |
| $\theta_Z$ | Zenith Angle (*degrees*) |
| $\theta_F$ | Angle of incidence at the front face of the panel |
| Pitch (p) | Row-to-row distance between the bottom edges of consecutive arrays (m) |
| Height (h) | Height of the panel |
| E | Elevation |
| $\beta$ | Tilt angle |
| $\gamma_A$ | Azimuth angle from measure from the North |
| $R_A$ | Albedo |
| C/ℂ | Cost / Cost per unit meter |
| M | Number of rows/arrays of modules |
| Z | Number of modules in an array |
| D | Yearly degradation rate in energy conversion |
| Y | Lifetime of a farm (in years) |
| YY | Yearly Yield |
| R | Discount rate |
| F | Fitting parameter in Alam-Sun model which effectively accounts for module assembly and its related effects (like transmittance, glazing, heat loss coefficient) |

| Sub/Super-scripts | Definition |
|---|---|
| a | Ambient |
| Alb | Albedo |
| bos | Balance of system |
| bot | Back side |
| DHI | Diffuse Horizontal Irradiance |
| diff | Diffuse |
| dir | Direct |
| dl | Differential element along the height of a panel |
| DNI | Direct Normal Irradiance |
| F | Front |
| f | Fixed cost |
| Farm | PV Farm |
| GHI | Global Horizontal Irradiance |
| Gnd | Ground |
| in | Input |
| l/L | Land |
| M, cell | cell/module |
| om | Operation and maintenance |
| Panel | PV panel |
| PV | Photovoltaic |
| rv | Residual value |
| STC | Standard testing condition |
| sys | System |
| top | top side |
| Total | Total |
| Z | Zenith |